\documentclass[11pt,a4paper]{article}

\usepackage[margin=2.3cm]{geometry}
\usepackage[T1]{fontenc}
\usepackage[utf8]{inputenc}
\usepackage{lmodern}
\usepackage{microtype}
\usepackage{graphicx}
\usepackage{booktabs}
\usepackage{amsmath,amssymb}
\usepackage{siunitx}
\usepackage{xcolor}
\usepackage{enumitem}
\usepackage[backend=biber,style=numeric,sorting=none]{biblatex}
\AtBeginBibliography{\small}
\usepackage[hidelinks]{hyperref}
\usepackage{caption}
\usepackage{subcaption}
\usepackage{float}

\newcommand{\lFEG}{\lambda_{\mathrm{FEG}}}
\newcommand{\RFEG}{R^2_{\mathrm{FEG}}}

\title{Data-Driven Characterisation of Wave-Forced Turbulence Using Time-Resolved Forecast-Error Growth\\ }

\author{
Raj Jyoti Baishya$^{1}$ \and
Joychen Kenglang$^{1}$ \and
Andrei Velichko$^{2}$ \and
Bimlesh Kumar$^{1,*}$\\[0.5em]
\small $^{1}$Department of Civil Engineering, Indian Institute of Technology Guwahati, Guwahati, India\\
\small $^{2}$Institute of Physics and Technology, Petrozavodsk State University, Petrozavodsk 185910, Russia\\
\small $^{*}$Corresponding author: \texttt{bimk@iitg.ac.in}
}
\date{}

\begin{document}
\maketitle

\begin{abstract}
Periodic surface-wave forcing can reorganise turbulent flows through coherent spectral response, synchronization, intermittency, and changes in short-term predictability, so its dynamical effect need not vary monotonically with forcing frequency. We reanalyse laboratory acoustic Doppler velocimetry records obtained at constant discharge under four conditions (0, 0.5, 0.67, and 1~Hz) using one common KNN--GMAE forecast-error-growth protocol. A distance-weighted $k$-nearest-neighbour predictor generates out-of-sample forecasts over multiple horizons, and the slope of the early quasi-linear region of $\ln(\mathrm{GMAE})$ versus physical forecast time is reported as a finite-horizon forecast-error-growth rate, $\lFEG$. For the full 120-s records, $\lFEG$ is 5.68, 0.85, 3.39, and 4.11~s$^{-1}$ for 0, 0.5, 0.67, and 1~Hz, respectively. The ordering 0~Hz $>$ 1~Hz $>$ 0.67~Hz $>$ 0.5~Hz is preserved in all seven nearby parameter configurations, indicating that the comparative result is not an artefact of a single KNN setting. A 40-s sliding-window analysis with a 10-s step reveals substantial temporal structure. The no-wave condition remains predominantly high and the 0.5-Hz condition predominantly low, whereas the 1-Hz record has weaker local support for a single exponential-growth regime: only 3 of 9 windows satisfy the adopted early-fit criterion $\RFEG\geq0.90$, compared with 8/9, 7/9, and 7/9 for 0, 0.5, and 0.67~Hz. Here, $\RFEG$ is the coefficient of determination of the linear fit to the early $\ln(\mathrm{GMAE})$ growth interval and is not a KNN prediction-score metric. Normalised power spectra retain distinct forcing-related peaks in the wave-driven records. These results support a frequency-selective, time-dependent reorganisation of finite-horizon instability rather than a simple monotonic reduction of chaos with increasing wave frequency, and they show why local fit diagnostics are essential when Lyapunov-type quantities are inferred from experimental turbulent signals.
\end{abstract}


\section{Introduction}
Coastal and open-channel flows are frequently subjected to the combined action of currents and surface waves \cite{fredsoe1992,jonsson1990,soulsby1997,grant1979,nielsen1992}. The resulting wave--current interaction modifies near-bed momentum exchange, sediment mobility, bank and shoreline stability, and the loading environment of hydraulic structures. Much of the classical literature therefore focuses on mean-flow properties, wave boundary layers, Reynolds stresses, sediment transport, and energy dissipation. These quantities remain central, but recent laboratory and field studies increasingly show that wave forcing also reorganises the temporal and spatial structure of turbulence itself.

Experiments in smooth-bed open-channel flow have shown that collinear surface waves can weaken very-large-scale turbulent motions and introduce new spectral organisation in wave-dominated regions \cite{peruzzi2021}. Nearshore ADV measurements likewise reveal phase-dependent modulation of turbulent events under combined wave--current conditions \cite{marino2024}. Recent open-channel studies with periodic surface forcing or submergence further demonstrate systematic redistribution of turbulent energy and coherent structures across scales \cite{liu2024,druault2026}. Taken together, these results indicate that periodic wave action should not be viewed merely as an additive oscillatory component superimposed on otherwise unchanged turbulence; it can alter the organisation, intermittency, and scale coupling of the measured velocity fluctuations.

A broader literature on periodically forced fluid systems reinforces this point. External harmonic forcing can produce synchronization, desynchronization, subharmonic responses, quasiperiodicity, intermittency, and transitions to low-dimensional chaos depending on the relationship between forcing and intrinsic flow time scales \cite{herrmann2020,khodkar2020,yang2024}. Cyclostationary spectral analysis of harmonically forced turbulent flows also shows that forcing can restructure energetic modes and their harmonic interactions rather than simply change a mean intensity \cite{heidt2024}. In other forced-fluid settings, the response to a forcing-frequency ratio can be non-monotonic \cite{singh2024}. Consequently, there is no general dynamical reason to assume in advance that increasing wave frequency must monotonically increase or decrease chaotic intensity or predictability.

Chaos theory offers a natural framework for quantifying sensitivity to initial conditions. The largest Lyapunov exponent (LLE) characterises the exponential separation of nearby trajectories and is commonly related to the rate at which predictability is lost \cite{wolf1985,rosenstein1993,strogatz2018}. Classical scalar-time-series estimators remain important references, but estimates from experimental records can be sensitive to noise, record length, embedding choices, and the interval selected for fitting the divergence law \cite{kantz2004,balasuriya2020,brari2022}. Recent work has therefore explored supervised machine learning for local Lyapunov exponents \cite{ayers2023}, uncertainty-aware alternatives \cite{garcia2026}, and predictor-based formulations in which out-of-sample forecast errors provide a model-free proxy for trajectory separation.

A forecast-error-growth estimator based on the geometric mean absolute forecast error (GMAE) was recently validated against reference positive LLEs in canonical nonlinear maps \cite{velichko2025}. Its central idea is to infer the dominant divergence rate from the early quasi-linear slope of $\ln(\mathrm{GMAE})$ across forecast horizons. A subsequent profile-based extension, FEG-Pro, emphasises that finite records should not be reduced to a slope alone: fit regime, curvature, residual roughness, monotonicity, and temporal stability can reveal when a single exponential interpretation is poorly supported \cite{velichko2026fegpro}. This distinction is especially relevant for experimental turbulence, where the governing state is high-dimensional and only a scalar observable may be available.

Despite rapid progress in both wave--current turbulence measurements and data-driven instability analysis, direct scalar Lyapunov-type characterisation of experimentally measured wave-forced turbulent velocity records remains uncommon. Existing wave--current studies largely quantify turbulence through statistics, spectra, coherent structures, or sediment-related quantities, whereas contemporary Lyapunov-estimation studies are still dominated by canonical chaotic systems, geophysical models, or non-fluid experimental signals. The combination of controlled wave-frequency variation, ADV velocity measurements, and time-resolved forecast-error-growth analysis therefore addresses a current methodological and physical gap.

The present study revisits laboratory velocity records obtained at a constant discharge under four forcing conditions: no imposed wave and monochromatic waves at 0.5, 0.67, and 1~Hz. Rather than imposing a monotonic frequency hypothesis, we ask three related questions: whether the four conditions show reproducibly different finite-horizon forecast-error divergence under one common analysis protocol; whether the comparative ordering survives nearby choices of KNN representation parameters; and whether a single full-record slope is representative of the local behaviour throughout each 120-s record. A normalised power-spectral analysis is used as a complementary description of coherent forcing-related temporal structure.

\section{Materials and Methods}
\subsection{Experimental data and preprocessing}
\begin{figure}[H]
\centering
\includegraphics[width=0.98\linewidth]{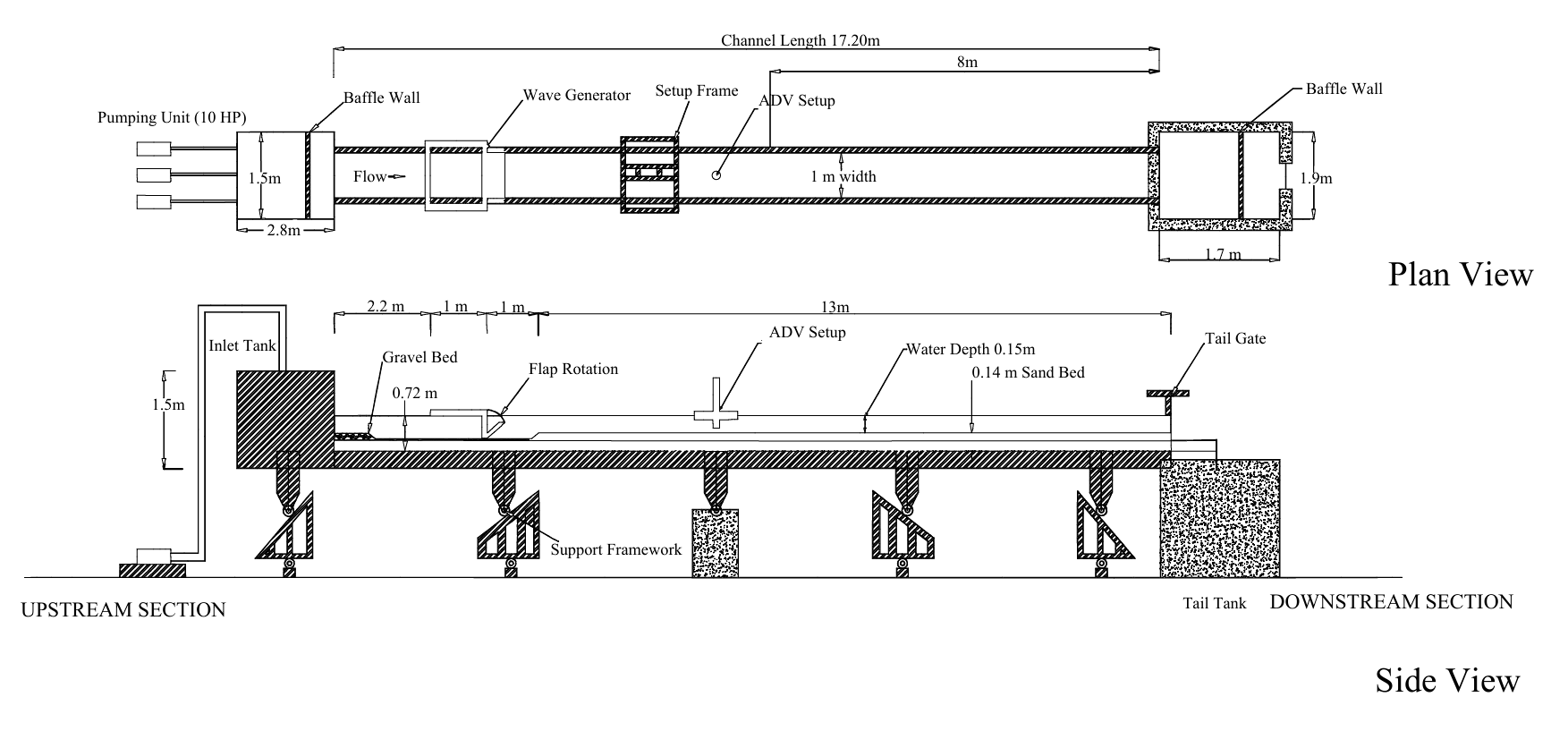}
\caption{Schematic diagram of the experimental setup.}
\label{fig:flume}
\end{figure}

The experiment was conducted at a constant discharge of 30~L/s. Figure~\ref{fig:flume} shows the experimental setup.
The main flume was 17.2~m long, 1.0~m wide, and 0.72~m deep. An upstream supply tank (2.8~m $\times$ 1.5~m $\times$ 1.5~m) fed the flume through a regulated inlet supplied by three 10-HP pumps. The discharge was adjusted with a control valve and measured using a rectangular weir in the downstream collection tank. A flow depth of 0.15~m was maintained using a downstream tailgate. The channel contained a 0.14-m sand bed whose upper surface was levelled before the measurements. A flap-type wave generator at the upstream end produced monochromatic forcing at 0.5, 0.67, and 1~Hz; the 0-Hz condition represents the baseline turbulent flow without imposed waves. Instantaneous velocity was measured with an acoustic Doppler velocimeter (ADV) at a fixed location. The sampling frequency was 100~Hz ($\Delta t=0.01$~s). Approximately 12,000 samples were retained for each condition, corresponding to about 120~s of data.

\begin{table}[H]
\centering
\footnotesize
\color{black}
\caption{Experimental wave and ADV parameters.}
\label{tab:experimental_parameters}
\begin{tabular}{lcccc}
\toprule
Condition & Wave height $H$ (m) & Wavelength $L$ (m) & ADV location $x$ (m) & ADV height above bed $z_b$ (m)\\
\midrule
0 Hz    & -- & -- & \textbf{5.7} & \textbf{0.0771}\\
0.5 Hz  & \textbf{0.034813} & \textbf{2.707} & \textbf{5.7} & \textbf{0.0786}\\
0.67 Hz & \textbf{0.037006} & \textbf{1.964} & \textbf{5.7} & \textbf{0.0785}\\
1 Hz    & \textbf{0.054106} & \textbf{1.212} & \textbf{5.7} & \textbf{0.0779}\\
\bottomrule
\end{tabular}
\color{black}
\end{table}

Following the experimental preprocessing protocol, the measured velocity was decomposed as
\begin{equation}
 u(t)=\overline{U}+u_w(t)+u'(t),
\end{equation}
where $\overline{U}$ is the mean-flow contribution, $u_w(t)$ is the wave-orbital component, and $u'(t)$ denotes the turbulent velocity fluctuation. The wave-orbital velocity was estimated from linear wave theory and removed from the measured signal after mean subtraction \cite{soulsby2006,dean1991dalrymple}. The horizontal wave-orbital velocity was represented as
\begin{equation}
u_w(x,z,t)=\frac{H}{2}\omega\frac{\cosh[k(h+z)]}{\sinh(kh)}\cos(kx-\omega t),
\end{equation}
where $H$ is the wave height, $h$ is the water depth, $z$ is the vertical coordinate measured upward from the still-water level ($z=0$ at the mean free surface and $z=-h$ at the bed), $x$ is the horizontal measurement location used in the wave-phase calculation, $k$ is the wave number, and $\omega$ is the angular frequency. If the ADV measurement volume is located at a height $z_b$ above the bed, then $z=z_b-h$. The wave number and angular frequency were calculated as $k=2\pi/L$ and $\omega=2\pi f$, respectively, where $L$ is the measured wavelength and $f$ is the forcing frequency. The processed turbulent component is therefore
\begin{equation}
 u'(t)=u(t)-\overline{U}-u_w(t).
\end{equation}
The original preprocessing also included ADV spike removal and noise correction consistent with established ADV-processing procedures \cite{goring2002,doroudian2010}. The four supplied analysis files contain the processed scalar time series in a column labelled \texttt{module}; this scalar series is denoted by $x(t)$ in the forecasting analysis below and is used identically for all four experimental conditions.

\subsection{Forecast-error-growth estimator}
The analysis follows the machine-learning forecast-error-growth framework of \textcite{velichko2025} and adopts the finite-horizon profile interpretation emphasised in FEG-Pro \cite{velichko2026fegpro}. Let a predictor generate an out-of-sample forecast $\hat{x}_{i+h}$ for a future value $x_{i+h}$ at horizon $h$. The absolute forecast error is
\begin{equation}
 e_i(h)=\left|x_{i+h}-\hat{x}_{i+h}\right|.
\end{equation}
For each horizon, the errors over the test set are aggregated by the geometric mean absolute error,
\begin{equation}
 \mathrm{GMAE}(h)=\left(\prod_{i=1}^{n} e_i(h)\right)^{1/n}
 =\exp\left[\frac{1}{n}\sum_{i=1}^{n}\ln e_i(h)\right],
\end{equation}
with a small numerical floor used only to avoid taking the logarithm of zero. If the early forecast-error growth is approximately exponential,
\begin{equation}
 \mathrm{GMAE}(t_h)\approx A\exp(\lFEG t_h),
\end{equation}
then
\begin{equation}
 \ln \mathrm{GMAE}(t_h)=\ln A+\lFEG t_h.
\end{equation}
The slope $\lFEG$ has units of s$^{-1}$. When a sufficiently supported quasi-linear early-growth regime exists, it is interpreted as a finite-horizon data-driven estimate of the dominant positive Lyapunov divergence rate. In the present experimental application we retain the notation $\lFEG$ to distinguish the measured finite-horizon forecast-error slope from an asymptotic exponent obtained from known governing equations. This conservative notation is also motivated by recent work showing that noisy and finite records can support useful local instability diagnostics even when an asymptotic Lyapunov interpretation is uncertain \cite{balasuriya2020,brari2022,velichko2026fegpro}.

\subsection{KNN state representation and prediction protocol}
A single analysis protocol was fixed before comparing the four experimental conditions. Three parameters define the KNN representation: the history span $H$, the embedding dimension $m$, and the KNN neighbour count $k$. The primary values are
\begin{equation}
 (H,m,k)=(450,30,3).
\end{equation}
At the 100-Hz sampling rate, $H=450$ samples corresponds to a physical history span of 4.5~s. The parameter $m=30$ is the number of historical signal values used as KNN input features; importantly, these are not 30 consecutive samples. Instead, integer offsets $\tau_j$ are distributed approximately uniformly over the complete interval $0\leq\tau_j\leq H-1$, and the state vector is
\begin{equation}
 \mathbf{z}_i=\left[x_{i+\tau_1},x_{i+\tau_2},\ldots,x_{i+\tau_m}\right],
 \qquad m=30.
\end{equation}
Thus, the 30-dimensional input vector represents the complete 4.5-s history while avoiding a 450-dimensional KNN search. In the primary configuration the offsets span samples 0 through 449. The parameter $k=3$ denotes the number of nearest neighbours used by the distance-weighted KNN regressor and is distinct from the embedding dimension $m$.

For each forecast horizon $h=1,\ldots,15$ samples (0.01--0.15~s), a separate supervised regression problem is constructed. The target is the value exactly $h$ samples after the last sample represented by the history window. Samples are divided chronologically into 60\% training and 40\% testing subsets, preserving temporal order. Feature means and standard deviations are estimated from the training subset only and used to standardise both training and test inputs. Prediction is then performed with Euclidean-distance KNN regression using distance weighting and $k=3$ neighbours. The complete procedure is repeated independently for every forecast horizon, and the test-set errors are aggregated by GMAE.

\subsection{Selection and quality of the early FEG slope}
The FEG slope is estimated only from the early part of the $\ln(\mathrm{GMAE})$ curve, before longer-horizon curvature or saturation can dominate. Candidate fits use the first $q$ forecast points with $q=5,6,\ldots,10$. For every candidate, a straight line is fitted to $\ln[\mathrm{GMAE}(t_h)]$ versus physical forecast time $t_h=h\Delta t$. A candidate is considered supported when
\begin{equation}
 \RFEG\geq0.90
 \quad\text{and}\quad
 \mathrm{nMAE}_{\mathrm{fit}}\leq0.12,
\end{equation}
where $\RFEG$ is the coefficient of determination of this early FEG linear regression and $\mathrm{nMAE}_{\mathrm{fit}}$ is the mean absolute fit residual normalised by the range of the fitted $\ln(\mathrm{GMAE})$ values. Among candidates satisfying both thresholds, the selected window maximises $\RFEG$, with lower $\mathrm{nMAE}_{\mathrm{fit}}$ and then a larger number of points used as tie-breakers. If no candidate passes the thresholds, the best early candidate is retained only as a diagnostic estimate and the window is explicitly flagged as a failed fit-quality case.

The notation $\RFEG$ is used throughout this paper to avoid ambiguity: it quantifies how well a straight line represents the selected early forecast-error-growth interval. It is \emph{not} the predictive $R^2$ of the KNN regression model. The KNN prediction errors enter the analysis through GMAE; $\RFEG$ is evaluated only after the multi-horizon GMAE curve has been constructed. The slope of the selected line is $\lFEG$ in s$^{-1}$. The same model parameters, fit-selection rules, and thresholds are applied to all four experimental conditions.

\subsection{Parameter-robustness analysis}
Because the absolute slope can depend on the representation and KNN parameters, a one-factor-at-a-time sensitivity analysis was performed around the primary protocol. Seven configurations were evaluated: the primary setting $(H,m,k)=(450,30,3)$; history spans $H=300$ and 600 with $m=30$, $k=3$; embedding dimensions $m=20$ and 40 (i.e., the number of historical feature coordinates distributed over the fixed $H=450$ span) with $k=3$; and neighbour counts $k=2$ and 5 with $H=450$, $m=30$. This analysis is used only to test whether the scientific ordering of the four conditions is robust; parameters are not optimised separately for individual records.

\subsection{Sliding-window analysis}
To resolve time dependence that is hidden by a single full-record slope, the same primary protocol was applied within overlapping 40-s windows shifted by 10~s. Nine windows were obtained from each 120-s record, with centres at approximately 20, 30, \ldots, 100~s. For every window we report the local $\lFEG$ and $\RFEG$ of the selected early linear interval.

Because adjacent windows overlap, these local estimates are not statistically independent experimental replicates. They are used as descriptive diagnostics of temporal variation rather than as independent samples for hypothesis testing.

\subsection{Spectral analysis}
A Welch-type power spectral density (PSD) estimate was calculated using Hann-windowed segments of 4096 samples with 50\% overlap. The spectra are displayed over 0--5~Hz. For comparison of spectral structure across conditions, each PSD is also normalised by its maximum value. The frequency resolution is approximately 0.0244~Hz.

\section{Results}
\subsection{Full-record forecast-error growth}
\begin{figure}[H]
\centering
\includegraphics[width=0.98\linewidth]{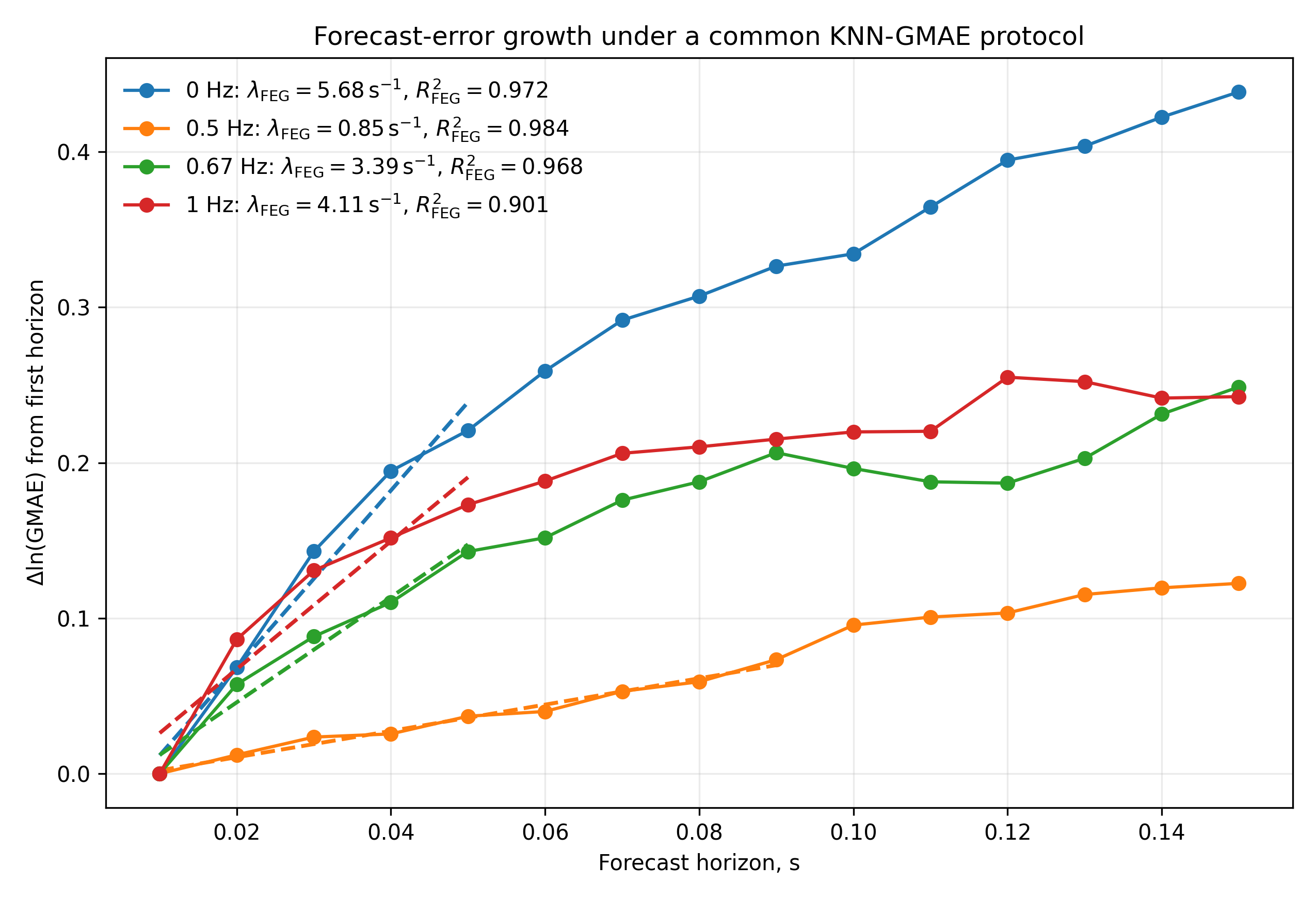}
\caption{Forecast-error-growth profiles under the common KNN--GMAE protocol. Curves are vertically aligned by subtracting the first $\ln(\mathrm{GMAE})$ value, which leaves the slopes unchanged. Dashed lines indicate the selected early fitting intervals. Each dashed fit is drawn in the same colour as its corresponding FEG curve. The legend reports the primary $\lFEG$ estimate and the corresponding early-fit $\RFEG$.}
\label{fig:feg}
\end{figure}

Figure~\ref{fig:feg} compares the four forecast-error-growth profiles after subtracting the first $\ln(\mathrm{GMAE})$ value from each curve. This vertical alignment changes only the intercept and therefore does not change the estimated slope. The no-wave condition exhibits the steepest early growth, whereas the 0.5-Hz condition exhibits a much shallower and nearly linear increase. The 0.67- and 1-Hz conditions are intermediate.

The primary full-record estimates are summarised in Table~\ref{tab:primary}. The resulting ordering is
\begin{equation}
 0~\mathrm{Hz} > 1~\mathrm{Hz} > 0.67~\mathrm{Hz} > 0.5~\mathrm{Hz}.
\end{equation}
Thus, the present reanalysis does not support a monotonic decrease of the divergence rate with increasing forcing frequency. In particular, the 0.5-Hz condition produces the smallest forecast-error-growth slope, while the 1-Hz condition is larger than the 0.67-Hz condition in the full-record calculation.

\begin{table}[H]
\centering
\caption{Primary full-record forecast-error-growth estimates.}
\label{tab:primary}
\begin{tabular}{lrrrrl}
\toprule
Condition & $\lFEG$ (s$^{-1}$) & $1/\lFEG$ (s) & Fit points & $\RFEG$ & Reliability\\
\midrule
0 Hz    & 5.679 & 0.176 & 5 & 0.972 & High\\
0.5 Hz  & 0.846 & 1.182 & 9 & 0.984 & High\\
0.67 Hz & 3.386 & 0.295 & 5 & 0.968 & High\\
1 Hz    & 4.114 & 0.243 & 5 & 0.901 & Medium\\
\bottomrule
\end{tabular}
\end{table}

The reciprocal $1/\lFEG$ is reported only as a characteristic e-folding time, not as a universal forecast horizon. A practical predictability horizon would additionally depend on the initial error level and the admissible forecast-error threshold.

\subsection{Robustness to analysis parameters}
\begin{figure}[H]
\centering
\includegraphics[width=0.92\linewidth]{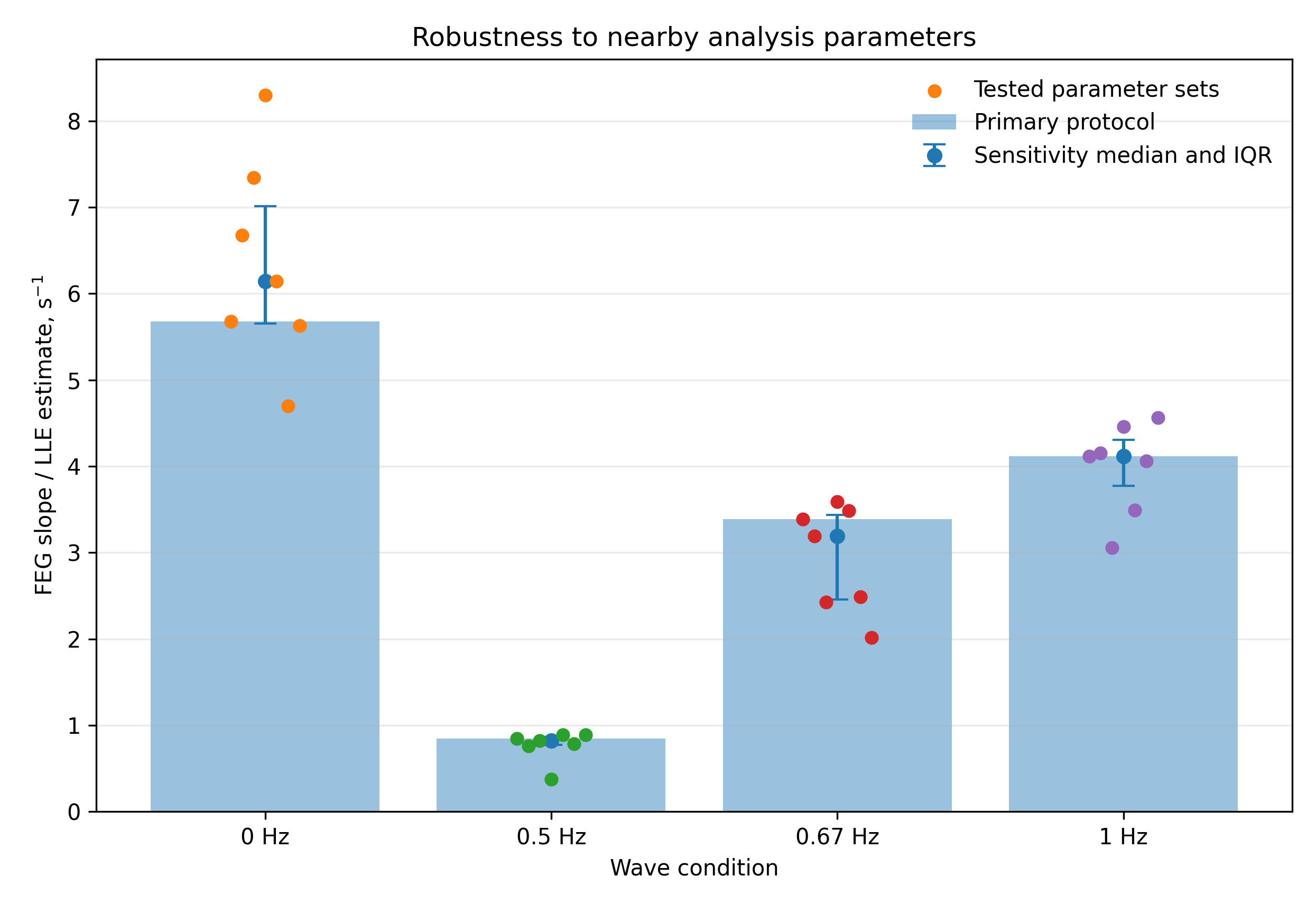}
\caption{Robustness of $\lFEG$ to nearby analysis parameters. Bars show the primary protocol, individual points show the seven tested parameter sets, and markers with error bars show the median and interquartile range. The ordering of all four wave conditions is unchanged in all seven configurations.}
\label{fig:robust}
\end{figure}

Figure~\ref{fig:robust} shows the primary estimate together with all seven tested parameter configurations and the median/interquartile range across configurations. Absolute values shift as the history span, embedding dimension, and neighbour count are varied, as expected for a finite-data estimator. However, the complete ordering 0~Hz $>$ 1~Hz $>$ 0.67~Hz $>$ 0.5~Hz is preserved in all 7 of 7 tested configurations. The median slopes over the seven configurations are 6.14, 0.82, 3.19, and 4.11~s$^{-1}$ for 0, 0.5, 0.67, and 1~Hz, respectively.

This robustness test is central to the interpretation: the scientific comparison does not rely on tuning a separate KNN configuration for each wave condition.

\subsection{Time-resolved sliding-window behaviour}
\begin{figure}[H]
\centering
\includegraphics[width=0.96\linewidth]{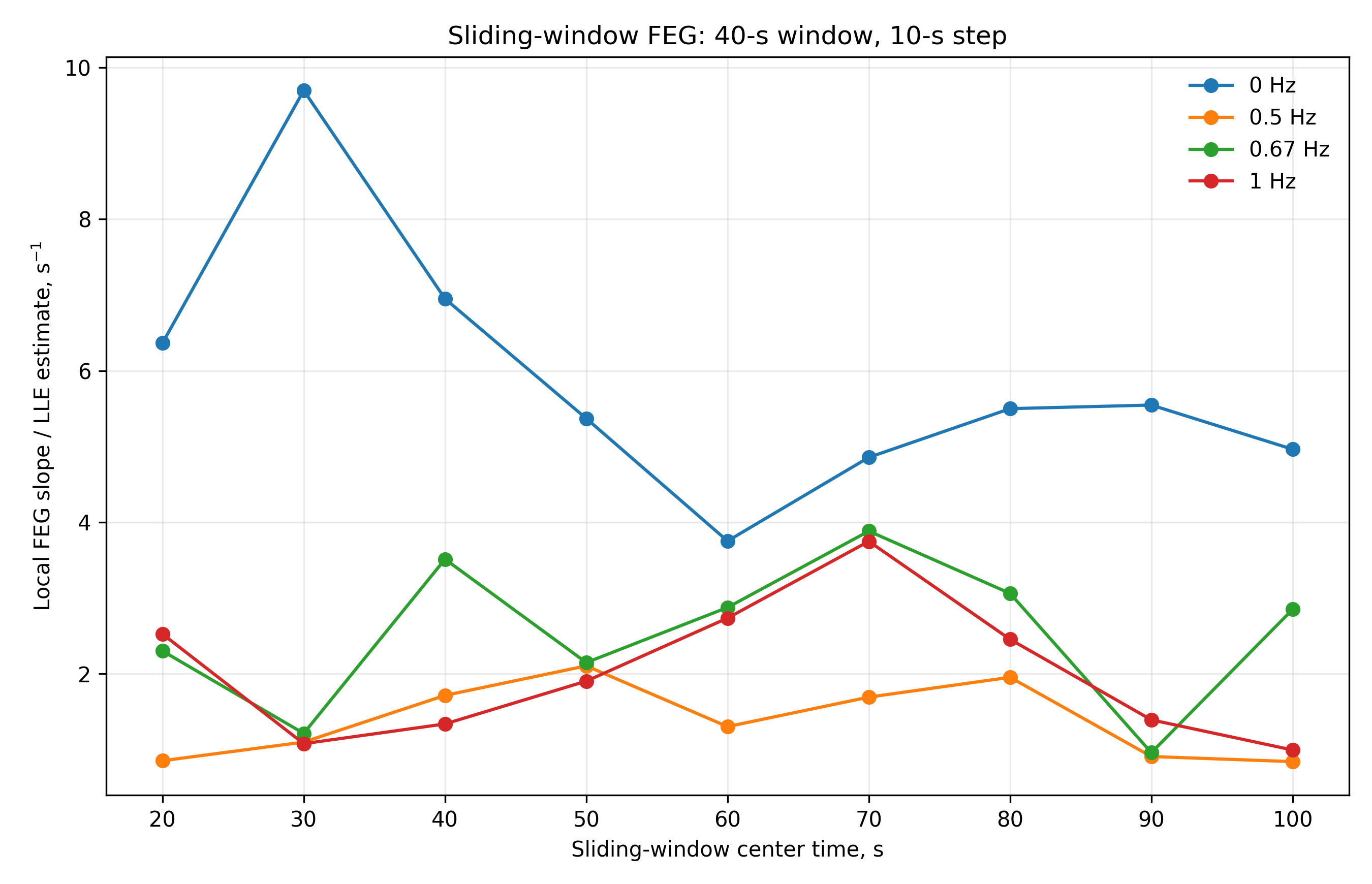}
\caption{Time-resolved local forecast-error-growth slope from overlapping 40-s windows shifted by 10~s. The no-wave condition remains predominantly high, whereas the 0.5-Hz condition remains low. The 0.67- and 1-Hz conditions show stronger temporal variation.}
\label{fig:sliding}
\end{figure}

The sliding-window analysis reveals structure that is not visible in the global 120-s estimates (Fig.~\ref{fig:sliding}). The no-wave condition remains predominantly high throughout the record and has a median local slope of 5.50~s$^{-1}$. The 0.5-Hz condition remains the low-divergence case, with a median of 1.30~s$^{-1}$ and values between approximately 0.84 and 2.11~s$^{-1}$. The 0.67-Hz and 1-Hz signals exhibit stronger temporal modulation, with medians of 2.85 and 1.90~s$^{-1}$, respectively.

\begin{figure}[H]
\centering
\includegraphics[width=0.96\linewidth]{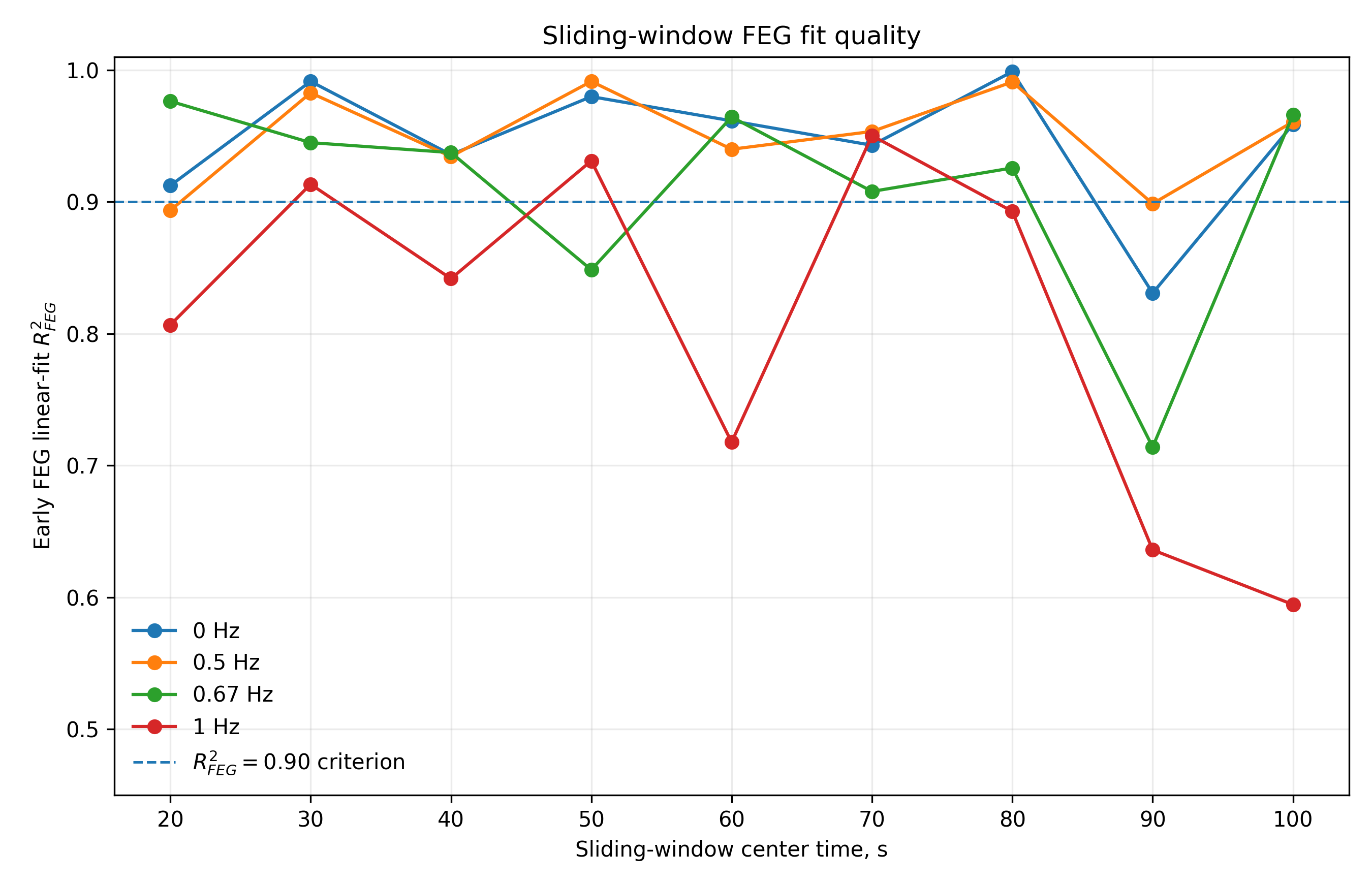}
\caption{Quality of the early linear fit for the sliding-window analysis. The dashed line marks the adopted $\RFEG=0.90$ criterion; $\RFEG$ refers to the early FEG linear fit, not to KNN predictive accuracy. Only 3 of 9 windows satisfy this criterion for the 1-Hz condition, compared with 8/9, 7/9, and 7/9 for 0, 0.5, and 0.67~Hz.}
\label{fig:slidingr2}
\end{figure}

The local fit quality provides an additional diagnostic (Fig.~\ref{fig:slidingr2}). For 0, 0.5, and 0.67~Hz, 8/9, 7/9, and 7/9 windows, respectively, satisfy the adopted $\RFEG\geq0.90$ criterion. In contrast, only 3/9 windows pass for 1~Hz, and the median $\RFEG$ is 0.842. The poor linearity in many 1-Hz windows indicates that a single global slope should be interpreted cautiously for this regime. In particular, the full-record value of 4.11~s$^{-1}$ reflects an aggregate behaviour that is not consistently reproduced by local windows.

\begin{table}[H]
\centering
\caption{Summary of the sliding-window analysis (40-s window, 10-s step).}
\label{tab:sliding}
\begin{tabular}{lrrrrr}
\toprule
Condition & Median $\lFEG$ & IQR (s$^{-1}$) & Range (s$^{-1}$) & Median $\RFEG$ & Fit pass\\
\midrule
0 Hz    & 5.499 & 4.963--6.364 & 3.753--9.696 & 0.959 & 8/9\\
0.5 Hz  & 1.304 & 0.909--1.716 & 0.842--2.105 & 0.953 & 7/9\\
0.67 Hz & 2.851 & 2.149--3.061 & 0.960--3.883 & 0.937 & 7/9\\
1 Hz    & 1.904 & 1.339--2.526 & 0.996--3.748 & 0.842 & 3/9\\
\bottomrule
\end{tabular}
\end{table}

\subsection{Spectral structure}
\begin{figure}[H]
\centering
\includegraphics[width=0.96\linewidth]{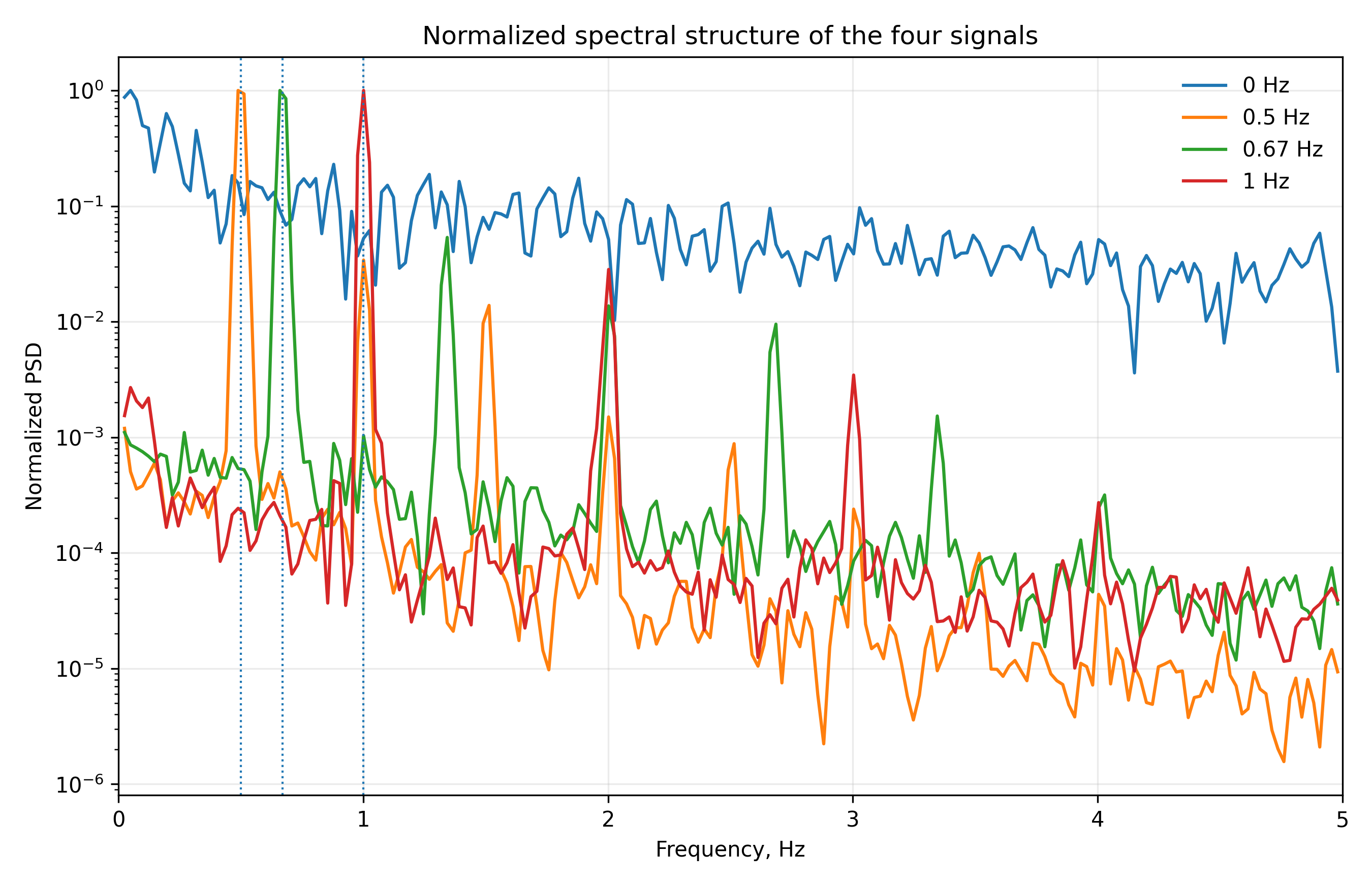}
\caption{Normalised power spectral density of the four analysed signals over 0--5~Hz. Dotted vertical lines mark the imposed forcing frequencies (0.5, 0.67, and 1~Hz). Strong peaks remain near the forcing frequencies in the wave-driven records.}
\label{fig:psd}
\end{figure}

The normalised PSDs are shown in Fig.~\ref{fig:psd}. Each forced record exhibits a strong narrow-band maximum close to the imposed wave frequency: approximately 0.488~Hz for the 0.5-Hz condition, 0.659~Hz for the 0.67-Hz condition, and 1.001~Hz for the 1-Hz condition. Harmonic structure is also visible. In contrast, the no-wave record exhibits broadband low-frequency content without a single forcing peak.

The coexistence of narrow-band forcing and broadband turbulent content is relevant to the forecast-error analysis. In particular, the low $\lFEG$ observed at 0.5~Hz is consistent with a signal containing a strong coherent, repeatedly predictable component. However, the spectral peak alone does not establish a causal explanation for the reduced forecast-error-growth rate.

\section{Discussion}
The reanalysis shows that periodic wave forcing changes short-term forecast-error divergence, but the dependence on forcing frequency is not monotonic. Under the common full-record protocol, the ordering is 0~Hz $>$ 1~Hz $>$ 0.67~Hz $>$ 0.5~Hz, and this complete ordering survives all seven nearby combinations of history span, embedding dimension, and KNN neighbourhood size. The absolute slopes vary with analysis parameters, as expected for finite experimental records, but the comparative structure is preserved. The strongest conclusion is therefore not the exact numerical value assigned to any one condition; it is the reproducible separation of the no-wave and 0.5-Hz regimes and the persistence of the full-record ranking under a controlled sensitivity test.

This non-monotonic response is compatible with the contemporary view of periodically forced fluid dynamics. Wave--current experiments show that surface waves can redistribute turbulent energy, weaken or reorganise large-scale motions, and modulate turbulent events in a phase-dependent manner \cite{peruzzi2021,marino2024,liu2024,druault2026}. More generally, harmonically forced wakes and jets exhibit synchronization boundaries, subharmonic responses, quasiperiodicity, intermittency, and transitions to chaos that depend on resonance and intrinsic system time scales rather than forcing frequency alone \cite{herrmann2020,khodkar2020,yang2024}. Cyclostationary SPOD results likewise demonstrate that periodic forcing can restructure energetic modes and harmonic couplings \cite{heidt2024}. Our data should therefore be interpreted as evidence for frequency-selective reorganisation of finite-horizon predictability, not as evidence for a universal law in which increasing wave frequency simply suppresses or enhances chaos.

The 0.5-Hz condition is particularly informative. It produces the smallest full-record $\lFEG$ and remains concentrated at comparatively low values in the sliding-window analysis. At the same time, its spectrum contains a pronounced narrow-band component near the imposed forcing frequency. A coherent recurrent component can make local analogue prediction easier and can therefore slow the growth of forecast error over short horizons. Such an interpretation is qualitatively consistent with synchronization and coherent-mode organisation reported in other periodically forced flows \cite{herrmann2020,heidt2024}. However, the PSD alone does not establish causality, and the present data do not demonstrate that the underlying high-dimensional turbulence has become globally less chaotic. The observable here is a finite-horizon instability rate extracted from one measured scalar signal.

The sliding-window results add information that a single 120-s estimate cannot provide. The no-wave signal remains predominantly high in local $\lFEG$, and the 0.5-Hz signal remains predominantly low, whereas the intermediate forcing cases show larger temporal rearrangement. The 1-Hz record is especially important because only 3 of 9 local windows satisfy the adopted $\RFEG\geq0.90$ criterion. Its full-record slope is therefore an aggregate descriptor of a signal for which a clean quasi-exponential early-growth region is not persistent in time. This is precisely the situation in which a profile-based interpretation is preferable to reporting one LLE-like number without diagnostics \cite{velichko2026fegpro}.

The local fit-quality result also places the methodology in the context of recent Lyapunov-estimation work. Classical and modern estimators alike are affected by finite data, observational noise, embedding or representation choices, and uncertainty in the inferred divergence law \cite{balasuriya2020,brari2022}. Supervised approaches can estimate local Lyapunov quantities when appropriate training information is available \cite{ayers2023}, while newer uncertainty-aware methods seek to characterise the instability of ensembles rather than rely on a single trajectory-separation construction \cite{garcia2026}. The present KNN--GMAE analysis occupies a different but complementary position: it requires no governing equations or tangent model and applies the same out-of-sample forecasting protocol to all experimental conditions. Its main safeguard is therefore transparent reporting of the early-fit quality and robustness to nearby parameter choices rather than an unsupported claim of asymptotic exactness.

A possible direction for future work is to treat persistent curvature of the multi-horizon forecast-error profile as dynamical information rather than only as a failed exponential fit. A recent preprint proposes comparing exponential forecast-error growth with Mittag--Leffler-type growth as a data-driven diagnostic for memory-compatible or fractional-like dynamics in scalar observations \cite{ngbo2026mittag}. The present data do not establish fractional-order turbulence, and no fractional order is estimated here. Nevertheless, the repeated loss of a clean early exponential-growth interval in parts of the 1-Hz record suggests a concrete future test: compare exponential and Mittag--Leffler-type descriptions of the same sliding-window GMAE curves, together with non-fractional curved alternatives. Even if a Mittag--Leffler model were preferred, its fitted order should initially be interpreted as an effective shape parameter of forecast-error growth rather than as direct identification of a governing fractional derivative order.

The spectral and forecast-error results together suggest that predictability in wave-forced turbulence reflects a competition between broadband turbulent variability and coherent periodically organised motion. Recent wave--current studies have shown that forcing changes spectral energy distribution, coherent structure topology, and phase-dependent turbulent transport \cite{peruzzi2021,marino2024,liu2024}. In the present records, the lowest full-record forecast-error growth occurs at 0.5~Hz rather than at the highest forcing frequency, and the 1-Hz record shows the weakest persistence of a locally linear growth regime. These observations argue against using forcing frequency itself as a surrogate for dynamical organisation. A more informative description combines frequency content with a time-resolved measure of how rapidly forecast errors diverge.

Several limitations delimit the claims that can be made. First, the sliding windows overlap and are descriptive local diagnostics, not independent experimental replicates; formal between-condition significance tests would require repeated experimental runs. Second, ADV records have finite sampling volume and measurement-noise limitations, which can affect high-frequency turbulence statistics and therefore any nonlinear time-series analysis based on the measured signal \cite{rastello2022}. Third, although the preprocessing protocol removes the theoretical wave-orbital component, the physical interpretation of forcing-related spectral peaks still depends on how completely coherent wave motion is separated from turbulent fluctuations. Finally, a scalar finite-horizon forecast-error slope should not be equated automatically with the asymptotic largest Lyapunov exponent of the full turbulent flow. These limitations do not remove the robust comparative findings, but they define the level at which they should be interpreted.

\section{Conclusions}
This study applies a single KNN--GMAE forecast-error-growth protocol to laboratory velocity records obtained under no-wave, 0.5-Hz, 0.67-Hz, and 1-Hz forcing and interprets the result as a finite-horizon instability profile rather than as an automatically exact asymptotic invariant. The full-record slopes follow the ordering 0~Hz $>$ 1~Hz $>$ 0.67~Hz $>$ 0.5~Hz, and that ordering remains unchanged in all seven nearby parameter configurations. The comparison is therefore robust to the tested variations of history span, embedding dimension, and KNN neighbour count. The no-wave record is consistently associated with the strongest short-horizon forecast-error divergence, while the 0.5-Hz condition is consistently associated with the weakest.

The time-resolved analysis shows why this comparison should not be reduced to four global numbers. Sliding 40-s windows reveal a persistent high-divergence tendency in the no-wave signal and a persistent low-divergence tendency at 0.5~Hz, but they also expose substantial temporal variation in the intermediate regimes. The 1-Hz condition is the clearest example: only three of nine windows support an early linear forecast-error-growth fit with $\RFEG\geq0.90$. For this regime, the full-record value is useful as an aggregate descriptor but is not representative of a stable local exponential-growth law throughout the experiment.

The spectral analysis provides a complementary view by showing narrow forcing-related peaks in the wave-driven signals against broadband turbulent content. Together with current literature on wave--current turbulence and periodically forced nonlinear flows, the results favour an interpretation in which forcing reorganises short-term predictability in a frequency-selective and time-dependent manner. They do not support the earlier simple picture of a monotonic reduction of chaotic intensity with increasing wave frequency. More generally, the study demonstrates that experimental Lyapunov-type analysis is strengthened when the estimated slope is accompanied by parameter-robustness checks, local fit quality, and temporal profiling.

Persistent departures from early exponential growth, especially in the 1-Hz sliding windows, also motivate future comparison with non-exponential forecast-error laws, including Mittag--Leffler-type profiles proposed as diagnostics of memory-compatible dynamics \cite{ngbo2026mittag}. This possibility is raised here only as a future direction and is not used to infer fractional-order dynamics from the present experiment.

\section*{Data and Code Availability}
The processed time-series data and the Python implementation used to reproduce the analyses are available from the corresponding authors upon reasonable request.

\section*{Author Contributions}
JK and RJB did the experiments and analyzed the observations. AV applied the methodology and interpreted the results. BK conceptualized the idea and supervised the work. JK and RJB wrote the initial draft. AV reviewed the draft and refined the original draft. All authors checked and refined the final draft.

\section*{Funding}
This research received no external funding. 

\section*{Conflict of Interest}
The authors declare no conflict of interest. 

\printbibliography

@book{fredsoe1992,
  author    = {Freds{\o}e, J. and Deigaard, R.},
  title     = {Mechanics of Coastal Sediment Transport},
  publisher = {World Scientific},
  year      = {1992}
}

@incollection{jonsson1990,
  author    = {Jonsson, I. G.},
  title     = {Wave--Current Interactions},
  booktitle = {The Sea},
  volume    = {9},
  publisher = {Wiley},
  year      = {1990}
}

@book{soulsby1997,
  author    = {Soulsby, R.},
  title     = {Dynamics of Marine Sands},
  publisher = {Thomas Telford},
  address   = {London},
  year      = {1997}
}

@article{grant1979,
  author  = {Grant, W. D. and Madsen, O. S.},
  title   = {Combined Wave and Current Interaction with a Rough Bottom},
  journal = {Journal of Geophysical Research},
  volume  = {84},
  pages   = {1797--1808},
  year    = {1979}
}

@book{nielsen1992,
  author    = {Nielsen, P.},
  title     = {Coastal Bottom Boundary Layers and Sediment Transport},
  publisher = {World Scientific},
  year      = {1992}
}

@book{strogatz2018,
  author    = {Strogatz, S. H.},
  title     = {Nonlinear Dynamics and Chaos},
  publisher = {CRC Press},
  year      = {2018}
}

@article{wolf1985,
  author  = {Wolf, A. and Swift, J. B. and Swinney, H. L. and Vastano, J. A.},
  title   = {Determining Lyapunov Exponents from a Time Series},
  journal = {Physica D},
  volume  = {16},
  pages   = {285--317},
  year    = {1985},
  doi     = {10.1016/0167-2789(85)90011-9}
}

@article{rosenstein1993,
  author  = {Rosenstein, M. T. and Collins, J. J. and De Luca, C. J.},
  title   = {A Practical Method for Calculating Largest Lyapunov Exponents from Small Data Sets},
  journal = {Physica D},
  volume  = {65},
  pages   = {117--134},
  year    = {1993},
  doi     = {10.1016/0167-2789(93)90009-P}
}

@book{kantz2004,
  author    = {Kantz, H. and Schreiber, T.},
  title     = {Nonlinear Time Series Analysis},
  publisher = {Cambridge University Press},
  year      = {2004}
}

@article{goring2002,
  author  = {Goring, D. G. and Nikora, V. I.},
  title   = {Despiking Acoustic Doppler Velocimeter Data},
  journal = {Journal of Hydraulic Engineering},
  volume  = {128},
  number  = {1},
  pages   = {117--126},
  year    = {2002}
}

@article{doroudian2010,
  author  = {Doroudian, B. and Nikora, V. and Goring, D.},
  title   = {On the Estimation of Turbulence Statistics from Acoustic Doppler Velocimeter Measurements},
  journal = {Journal of Hydraulic Engineering},
  volume  = {136},
  number  = {10},
  pages   = {719--732},
  year    = {2010}
}

@article{peruzzi2021,
  author  = {Peruzzi, C. and Vettori, D. and Poggi, D. and Blondeaux, P. and Ridolfi, L. and Manes, C.},
  title   = {On the Influence of Collinear Surface Waves on Turbulence in Smooth-Bed Open-Channel Flows},
  journal = {Journal of Fluid Mechanics},
  volume  = {924},
  year    = {2021},
  doi     = {10.1017/jfm.2021.605}
}

@article{marino2024,
  author  = {Marino, M. and Faraci, C. and Jensen, B. and Musumeci, R. E.},
  title   = {Turbulent Features of Nearshore Wave--Current Flow},
  journal = {Ocean Science},
  volume  = {20},
  pages   = {1479--1493},
  year    = {2024},
  doi     = {10.5194/os-20-1479-2024}
}

@article{liu2024,
  author  = {Liu, X. and Law, A. W.-K.},
  title   = {Turbulence Statistics and Structures in Fully Developed Open Channel Flows with Periodic Surface Coverages},
  journal = {Journal of Fluid Mechanics},
  volume  = {1000},
  pages   = {A6},
  year    = {2024},
  doi     = {10.1017/jfm.2024.921}
}

@article{druault2026,
  author  = {Druault, P. and Gaurier, B. and Germain, G.},
  title   = {{POD} Analysis of Regular Waves Interacting with Large Scale Flow Structures in the Water Column},
  journal = {Ocean Engineering},
  year    = {2026},
  doi     = {10.1016/j.oceaneng.2026.124601}
}

@article{herrmann2020,
  author  = {Herrmann, B. and Oswald, P. and Semaan, R. and Brunton, S. L.},
  title   = {Modeling Synchronization in Forced Turbulent Oscillator Flows},
  journal = {Communications Physics},
  volume  = {3},
  pages   = {195},
  year    = {2020},
  doi     = {10.1038/s42005-020-00466-3}
}

@article{khodkar2020,
  author  = {Khodkar, M. A. and Taira, K.},
  title   = {Phase-Synchronization Properties of Laminar Cylinder Wake for Periodic External Forcings},
  journal = {Journal of Fluid Mechanics},
  volume  = {904},
  year    = {2020},
  doi     = {10.1017/jfm.2020.772}
}

@article{yang2024,
  author  = {Yang, Z. and Guan, Y. and Redonnet, S. and Li, L. K. B.},
  title   = {Chaos via Type-{II} Intermittency in a Forced Globally Unstable Jet},
  journal = {Journal of Fluid Mechanics},
  volume  = {984},
  pages   = {R8},
  year    = {2024},
  doi     = {10.1017/jfm.2024.252}
}

@article{heidt2024,
  author  = {Heidt, L. and Colonius, T.},
  title   = {Spectral Proper Orthogonal Decomposition of Harmonically Forced Turbulent Flows},
  journal = {Journal of Fluid Mechanics},
  volume  = {985},
  pages   = {A42},
  year    = {2024},
  doi     = {10.1017/jfm.2024.70}
}

@article{singh2024,
  author  = {Singh, N. and Pal, A.},
  title   = {Quantifying the Turbulent Mixing Driven by the Faraday Instability in Rotating Miscible Fluids},
  journal = {Physics of Fluids},
  year    = {2024},
  doi     = {10.1063/5.0187973}
}

@article{ayers2023,
  author  = {Ayers, D. and Lau, J. and Amezcua, J. and Carrassi, A. and Ojha, V.},
  title   = {Supervised Machine Learning to Estimate Instabilities in Chaotic Systems: Estimation of Local Lyapunov Exponents},
  journal = {Quarterly Journal of the Royal Meteorological Society},
  volume  = {149},
  number  = {753},
  pages   = {1236--1262},
  year    = {2023},
  doi     = {10.1002/qj.4450}
}

@article{brari2022,
  author  = {Brari, Z. and Belghith, S.},
  title   = {A New Algorithm for Largest Lyapunov Exponent Determination for Noisy Chaotic Signal Studies with Application to Electroencephalographic Signals Analysis for Epilepsy and Epileptic Seizures Detection},
  journal = {Chaos, Solitons \& Fractals},
  volume  = {165},
  pages   = {112757},
  year    = {2022},
  doi     = {10.1016/j.chaos.2022.112757}
}

@article{balasuriya2020,
  author  = {Balasuriya, S.},
  title   = {Uncertainty in Finite-Time Lyapunov Exponent Computations},
  journal = {Journal of Computational Dynamics},
  volume  = {7},
  number  = {2},
  pages   = {313--337},
  year    = {2020},
  doi     = {10.3934/jcd.2020013}
}

@article{garcia2026,
  author  = {Garc{\'i}a-Guti{\'e}rrez, A. and Rubio, C. and Dom{\'i}nguez, D. and L{\'o}pez, D.},
  title   = {Chaos Meets Stochasticity: A Variance-Based Method for Lyapunov Exponent Estimation},
  journal = {Chaos},
  volume  = {36},
  number  = {1},
  pages   = {013126},
  year    = {2026},
  doi     = {10.1063/5.0311209}
}

@article{rastello2022,
  author  = {Rastello, M. and Klema, M. R. and Carpenter, A. B. and Garanaik, A. and Venayagamoorthy, S. K. and Gates, T. and Mari{\'e}, J.},
  title   = {Velocity Measurements in Developing Narrow Open-Channel Flows with High Free-Stream Turbulence: Acoustic Doppler Velocimetry ({ADV}) vs Laser Doppler Anemometry ({LDA})},
  journal = {Flow Measurement and Instrumentation},
  year    = {2022},
  doi     = {10.1016/j.flowmeasinst.2022.102206}
}

@article{velichko2025,
  author  = {Velichko, Andrei and Belyaev, Maksim and Boriskov, Petr},
  title   = {A Novel Approach for Estimating Largest Lyapunov Exponents in One-Dimensional Chaotic Time Series Using Machine Learning},
  journal = {Chaos},
  volume  = {35},
  pages   = {101101},
  year    = {2025},
  doi     = {10.1063/5.0289352}
}

@article{velichko2026fegpro,
  author        = {Velichko, Andrei and N'Gbo, N'Gbo and Carpentieri, Bruno and Shams, Mudassir},
  title         = {{FEG-Pro}: Forecast-Error Growth Profiling for Finite-Horizon Instability Analysis of Nonlinear Time Series},
  journal       = {arXiv preprint},
  year          = {2026},
  eprint        = {2605.17282},
  archiveprefix = {arXiv},
  primaryclass  = {nlin.CD},
  doi           = {10.48550/arXiv.2605.17282}
}

@article{ngbo2026mittag,
  author        = {N'Gbo, N'Gbo and Velichko, Andrei},
  title         = {Mittag-Leffler-Type Forecast-Error Growth as a Diagnostic Indicator of Fractional Dynamics},
  journal       = {arXiv preprint},
  year          = {2026},
  eprint        = {2607.08588},
  archiveprefix = {arXiv},
  primaryclass  = {math.DS},
  doi           = {10.48550/arXiv.2607.08588}
}

@book{dean1991dalrymple,
  author    = {Dean, Robert G. and Dalrymple, Robert A.},
  title     = {Water Wave Mechanics for Engineers and Scientists},
  publisher = {World Scientific},
  year      = {1991}
}

@techreport{soulsby2006,
  author      = {Soulsby, Richard L.},
  title       = {Simplified Calculation of Wave Orbital Velocities},
  institution = {HR Wallingford},
  number      = {TR 155},
  year        = {2006}
}

\end{document}